\documentclass[aps,prl,twocolumn,groupedaddress,showpacs]{revtex4-1}
\usepackage[english]{babel}
\usepackage{setspace}  
\usepackage{graphicx}
\usepackage{amssymb}
\usepackage{amscd}
\usepackage{rotating}
\usepackage{color}
\usepackage{epstopdf}
\begin{document}
\title{Percolation Enhanced Supersolids in the Extended Bose-Hubbard Model}

\author{B. M. Kemburi and V. W. Scarola}

\affiliation{Physics Department, Virginia Tech, Blacksburg, Virginia 24061, USA}

\begin{abstract}
We theoretically study the stability of lattice supersolid states in the extended Bose-Hubbard model with bounded spatial disorder.  We construct a disorder mean field theory and compare with quantum Monte Carlo calculations.  The supersolid survives weak disorder on the simple cubic lattice.  We also find that increasing disorder strength can transform a lattice solid into a supersolid as it tends to percolate through the disorder landscape.\end{abstract}
\maketitle

%%%%%%

Studies of the extended Bose-Hubbard (BH) model seek to capture the essential properties of a wide variety of physical systems including helium \cite{matsubara:1956,liu:1973}, Josephson junction arrays \cite{otterlo:1995}, and certain narrow-band superconductors \cite{micnas:1990}.  The model reveals a lattice supersolid phase (a simultaneous superfluid and solid) in its mean field phase diagram \cite{liu:1973,otterlo:1995}.  Recent experiments in solid $^{4}$He demonstrate evidence \cite{kim:2004} for the long-anticipated supersolid phase.  Disorder and defects are believed to impact the robustness of the observed phase \cite{clark:2007}.

Theoretical studies of the lattice supersolid suggest that it is rather delicate.  For example, quantum Monte Carlo (QMC) analyses reveal that large interaction strengths are required to stabilize the supersolid against quantum and thermal fluctuations in low dimensions \cite{scalettar:1995,batrouni:2000,sengupta:2005}.  More troubling results on the square lattice \cite{bernardet:2002} show that a different type of perturbation, external spatial disorder, destroy the solid itself leaving no chance for the supersolid.  The latter results stem from an Imry-Ma-type argument \cite{imry:1975,bernardet:2002} implying that the solid is unstable in the presence of arbitrarily weak disorder in less than three dimensions (3D).  Recent QMC results firmly established that in 3D, e.g., on the simple cubic (SC) lattice \cite{yamamoto:2009}, the supersolid is in fact stable against both thermal and quantum fluctuations yet, to the best of our knowledge, no QMC studies on the effect of spatial disorder on the SC lattice supersolid have been reported.  On the SC lattice we expect the stability of the solid against disorder \cite{sengupta:2005} but the fate of the supersolid is unknown.

Disorder in BH models generates an intriguing compressible glassy state, the Bose-glass \cite{fisher:1989}, that could compete with supersolids.  Considerable controversy (see Ref.~\cite{pollet:2009} and references therein) regarding the precise location of the Bose-glass in phase diagrams of the ordinary BH model (the short-range limit of the extended BH model) with disorder stems from Griffiths effects.  Griffiths effects arise from statistically rare, but relevant, disorder configurations that can generate glassy states.  These rare configurations combine with strong quantum fluctuations over large length scales.  As a result they are often missed in mean field theories (MFTs) and can mislead finite-sized QMC studies.  

The theorem of inclusions \cite{pollet:2009} establishes a key feature in the phase diagram topology of disordered BH models: We expect a compressible glassy state to completely surround incompressible states.  But in the extended BH model, supersolids are also expected to lie near the incompressible solid in parameter space.  The theorem of inclusions thus mandates a competition between glassy states and supersolids in extended BH models, opening the possibility that supersolids do not survive even weak disorder on cubic lattices.   

Work on the relationship between quantum order and disorder has become more pressing with the advent of cold atomic gas experiments demonstrating controlled disorder \cite{chen:2008,billy:2008,roati:2008,pasienski:2010}.  Disorder in optical lattice experiments, imposed with speckle laser light or bichromatic incommensuration, exhibit intriguing insulating phases of bosons \cite{roati:2008,billy:2008,pasienski:2010}.  Long range interactions among bosons, caused by dipolar moments \cite{goral:2002}, band effects \cite{scarola:2005}, or other mechanisms \cite{bloch:2008} suggest that the disordered extended BH model could underlie the essential properties of future optical lattice experiments.  

We examine the stability of disordered lattice supersolids in the presence of quantum and thermal fluctuations.  We construct a MFT phase diagram of the softcore extended BH model with disorder and compare with QMC.  For weak disorder we find that, in contrast to hardcore bosons on the square lattice \cite{bernardet:2002}, the solid and the supersolid are indeed stable on the SC lattice.  We then increase disorder to examine the role of strong disorder on bosonic solids.  We find a striking effect:  Disorder transforms the solid into a percolating supersolid in direct analogy to disorder triggered superfluidity via percolation in BH models \cite{krauth:1991,sheshadri:1995,dang:2009,pollet:2009}.  We construct a mean field site-percolation picture to qualitatively capture the disorder triggered supersolid behavior.

We study the interplay between disorder and supersolids with the extended BH model:
\begin{eqnarray}
H&=&-  t\sum_{\langle i,j \rangle} \left( b^{\dagger}_{i} b^{\phantom{\dagger}}_{j} + \text{H.c.}\right)+\frac{U}{2}\sum_{j} n_{j} (n_{j}-1) \nonumber \\
  &+&V\sum_{\langle i,j \rangle} n_{i}n_{j}-\sum_{j}\mu_{j}n_{j},
  \label{Hubbard}
\end{eqnarray}
where the operator $b^{\dagger}_{j}$ creates a boson at the lattice site $j$ and the number operator is $n_{j}=b^{\dagger}_{j} b^{\phantom{\dagger}}_{j} $.  The first term imposes a hopping energy gain ($t$) among nearest neighbors.  The second and third terms impose energy penalties for multiple occupancy at a site ($U$) and occupancy of nearest neighbor sites ($V$), respectively.  We consider a bipartite lattice with periodic boundaries.  To study a regime consistent with spatially decaying interactions and a strong supersolid we choose $zV=U=1$ \cite{sengupta:2005, yamamoto:2009}, where $z$ is the coordination number.  Our MFT applies to any bipartite lattice but when using QMC with Eq.~(\ref{Hubbard}) we will work on the SC lattice, $z=6$, with $L^{3}$ sites.  The last term denotes a spatially varying chemical potential: $\mu_{j}=\mu +\delta\mu_{j}$ with bounded disorder defined by the random number: $\delta\mu_{j}\in[-\Delta,\Delta]$.  

\begin{table}[t]
\caption{List of possible phases and related order parameters: static structure factor $S_{\pi}$, superfluid stiffness  $\rho_{s}$, and compressibility $\kappa$.} \centering
\begin{tabular}{|l|c|c|c|}
  \hline
              & $S_{\pi}$          & $\rho_{s} $         & $\kappa$   \\[-.1ex] \hline
  Solid     & $\neq0$  & $=0$ & $=0$  \\[-.1ex]\hline
  Superfluid      & =0 & $\neq 0$  & $\neq0$  \\\hline
  Supersolid     & $\neq0$ & $\neq0$  & $\neq0$ \\\hline
  Disordered solid	       & $\neq0$ & $=0$  & $\neq0$   \\\hline
  Bose glass	       & $=0$ & $=0$  & $\neq0$   \\
  \hline
\end{tabular}
\label{tab1}
\end{table}

We first examine the low-temperature phase diagram with a MFT that includes spatial disorder and thermal fluctuations but excludes quantum fluctuations.  We decouple sites by defining two density and two phase order parameters:     
$m_{\alpha}=\langle n_{\alpha} \rangle$ and $\phi_{\alpha}=\langle b^{\dagger}_{\alpha} \rangle$, respectively.    
$\alpha\in \{A,B\}$ denotes the sublattice index.  Here and in the following the expectation values include disorder averaging.  The mean field Hamiltonian \cite{yamamoto:2009} becomes:
$H^{\text{MF}}\equiv H_{A}+H_{B}+C$, where:
\begin{eqnarray}
H_{\alpha}&=&-zt(b^{\dagger}_{\alpha} +b^{\phantom{\dagger}}_{\alpha})\phi_{\alpha'}
+\frac{U}{2} n_{\alpha} (n_{\alpha}-1)
\nonumber\\
&+&zVn_{\alpha}m_{\alpha'}-\tilde{\mu}n_{\alpha},
\end{eqnarray}
and $C\equiv2zt\phi_{A}\phi_{B}-zVm_{A}m_{B}$.  In the above the prime denotes $\alpha\neq\alpha'$.  $\tilde{\mu}=\mu+\epsilon$ contains a continuously tunable parametrization of disorder, $\epsilon$.

Self-consistent minimization of the free energy with respect to the order parameters solves the mean field equations.  The mean field free energy is:
\begin{eqnarray}
F^{\text{MF}}&=&C-\beta^{-1}\ln Tr\left\{ e^{-\beta(H_{A}+H_{B})} \right\} \nonumber\\
&=&C-\beta^{-1}\ln \sum_{\alpha,\gamma} e^{-\beta E_{\alpha,\gamma}},
\end{eqnarray}
where $\beta$ is the inverse temperature and $E_{\alpha,\gamma}$ denotes the $\gamma^{\text{th}}$ energy eigenvalues of $H_{\alpha}$.  In the following we work at low densities.  We find that a restricted Fock number eigenstate basis accurately captures the low density properties: $\vert N \rangle=\{\vert 0\rangle,\vert 1\rangle,\vert 2\rangle\}$.  We then diagonalize the $3\times3$ matrix $\langle N \vert H_{\alpha}\vert N'\rangle$ to obtain three eigenvalues for each $\alpha$, $E_{\alpha,\gamma}$, where $\gamma=1,2,3$.

We use the free energy to solve the coupled mean field equations.  We numerically solve the following four integral equations:
 $I[\partial_{\phi_{\alpha}}F^{\text{MF}}]=0$ and $I[\partial_{m_{\alpha}}F^{\text{MF}}]=0$, for $m_{\alpha}$ and  $\phi_{\alpha}$.  Here $I[R]\equiv\int_{-\Delta}^{\Delta}d\epsilon (R/2\Delta)$ denotes an integral over bounded disorder.

The mean field equations yield several different ground states that are defined by combinations of disorder-averaged order parameters.  Solid order is defined by long-range oscillations in the density-density correlation function (diagonal long-range order in the density matrix) or, equivalently, peaks in the static structure factor at wavevector, ${\bf Q}$:
\begin{eqnarray}
S_{{\bf Q}}\equiv L^{-6}\sum_{j,k}e^{i{\bf Q}\cdot({\bf r}_{j}-{\bf r}_{k})}\langle n_{j} n_{k}\rangle, 
\end{eqnarray}
that indicate a spontaneous breaking of the sublattice symmetry.  For the large values of $V$ considered here an oscillation of the density between sublattices is favored, ${\bf Q}=(\pi,\pi,\pi)$ on the SC lattice.  In the mean field limit this corresponds to $S^{\text{MF}}_{\pi}\equiv(m_{A}-m_{B})^{2}/4$.

The superfluid phase is defined by off-diagonal long range order in the density matrix.  The superfluid stiffness is: 
\begin{eqnarray}
\rho_{s}\equiv L^{-6}\sum_{j,k}\langle b^{\dagger}_{j} b^{\phantom{\dagger}}_{k} + \text{H.c.} \rangle.
\label{stiffness}
\end{eqnarray}
This representation is well approximated by the stiffness measured in QMC, $\langle W^{2}\rangle/3t\beta$, where $W$ is the winding number \cite{pollock:1987,yamamoto:2009}.  In the mean field limit Eq.~(\ref{stiffness}) yields $\rho_{s}^{\text{MF}}\equiv(\phi_{A}+\phi_{B})^{2}/2$.  The supersolid is defined by coexisting superfluid and solid order.  The first three rows of Table~\ref{tab1} indicate expected orders in the uniform limit, $\Delta=0$.  

\begin{figure}[t] 
   \centering
   \includegraphics[width=3in]{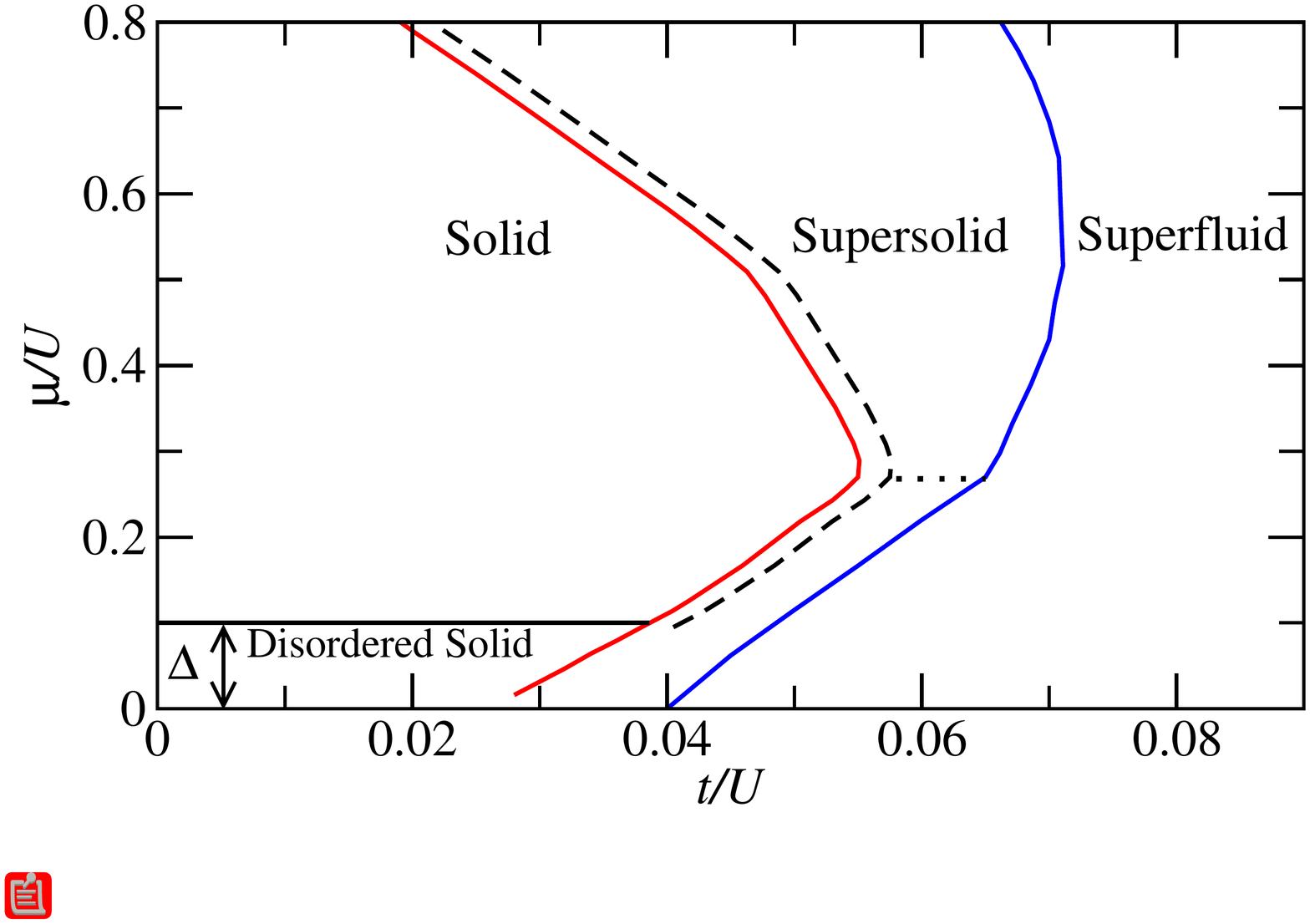} 
   \vspace{-1.5cm}
   \caption{(Color online) Solid lines show the zero temperature phase diagram of the extended Bose-Hubbard model, Eq.~(\ref{Hubbard}), with weak disorder, $\Delta/U=0.1$,  computed with MFT.  The strong nearest neighbor repulsion $zV/U=1$ opens a large solid lobe surrounded by the supersolid.  The phase diagram is similar to the non disordered case except for low $\mu$ (below the horizontal solid line) where the compressibility is finite.  The dotted line marks a possible transition between different supersolids. The dashed line is a schematic of the location of an expected quantum Griffiths-type region.} 
   \label{fig1}
\end{figure}

New phases arise in the presence of disorder, $\Delta> 0$.  For $V=U/z$ we expect the solid phase to dominate but at low $\mu$ the spatially varying part of the chemical potential can disrupt the perfect solid.  Here even weak disorder can create compressible domains that extend across the system.  We define a disordered solid as a solid which has a finite $S_{\pi}$, but zero global compressibility, $\kappa=\beta (\langle n^{2} \rangle -\langle n\rangle^{2})$ (Table~\ref{tab1}).

\begin{figure}[t] 
   \centering
   \includegraphics[width=3.2in]{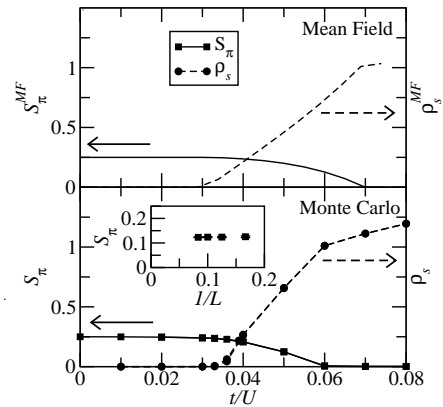} 
   \caption{The top panel shows the structure factor (solid line: left $y$-axis) and stiffness (dashed line: right $y$-axis) vs. hopping for $\mu/U=0.7$, $zV/U=1$, $\beta U=20$, and $\Delta/U=0.1$ computed using MFT.   The bottom panel shows the same as the top panel but computed with QMC on two different system sizes, $L=8$ and $10$.  Both system sizes yield nearly indistinguishable results.  The squares (circles) show results for the structure factor (stiffness) and the lines are guides to the eye.  The inset uses finite size scaling at $t/U=0.05$ to show that $L=8$ approximates the thermodynamic limit.  Both the top and bottom panels show that both methods support overlapping order parameters (i.e., a supersolid) for weak disorder.     }
   \label{fig2}
\end{figure}

We first use MFT to test the stability of the supersolid against weak disorder.   The MFT phase diagram (Fig.~\ref{fig1}) shows that the solid and the supersolid remain stable for intermediate chemical potentials $ \Delta <\mu<Vz-\Delta$. (The top panel of Fig.~\ref{fig2} shows MFT order parameters for a characteristic line in the phase diagram at non-zero temperature.)  But for $\mu<\Delta $ (or $\mu >Vz-\Delta$), disorder will tend to push the chemical potential beyond the gap of the uniform solid.  The energy gap (at $\Delta=0$) for the solid is given by the width of the solid lobe at fixed $t$, $E_{g}^{\text{S}}=\mu_{\text{max}}^{\text{S}}-\mu_{\text{min}}^{\text{S}}$, in the $\mu$ vs. $t$ phase diagram.  Here $\mu_{\text{max}}^{\text{S}}$ ($\mu_{\text{min}}^{\text{S}}$) is the largest (smallest) chemical potential allowing the solid phase at fixed $t$.  

Fig.~\ref{fig1} shows that for $\mu<\Delta$ the solid becomes compressible, $\kappa>0$.  The horizontal line located at $\mu=\Delta$ marks a transition from the solid to the disordered solid. (At finite temperatures this line becomes a crossover in MFT.)  The MFT is defined locally and therefore just measures a local compressibility.  In what follows we will describe QMC calculations.   Our preliminary QMC results show $\kappa>0$ for $\mu<\Delta$.  We tentatively conclude that the disordered solid appears for $\mu<\Delta$.

 The dotted line separates two different supersolid regions.  MFT shows a cusp in the stiffness along the dotted line.  Below the dotted line we expect a supersolid defined by combining a superfluid with a disordered solid that has compressible domains with a different spatial scaling behavior than above the dotted line.   Domain size scaling and transitions marked by horizontal lines will be explored in future work.  In the grand canonical ensemble phase separation (between a superfluid and a solid) manifests as a density discontinuity along phase boundaries \cite{batrouni:2000,sengupta:2005}.  Our preliminary QMC results indicate phase separation only near the dashed line for $t/U\sim0.04-0.05$.
  
We study the impact of quantum fluctuations using QMC.  Equation~(\ref{Hubbard}) does not have a sign problem and therefore allows numerically exact studies using QMC on finite sized systems.  We use the stochastic series expansion representation with directed loop updates \cite{syljuasen:2002} within the Algorithms and Libraries for Physics Simulations (ALPS) framework \cite{albuquerque:2008} to evaluate order parameters of Eq.~(\ref{Hubbard}).  We average over $\sim 500-1000$ disorder realizations for each data point reported.  For each disorder configuration a sufficient number of thermalized QMC steps are chosen to yield convergence of QMC error.  Error bars indicate the standard deviation in disorder averaging.

The bottom panel of Fig.~\ref{fig2} demonstrates the stability of the solid and the supersolid in the presence of quantum fluctuations and spatial disorder.  Two different system sizes are plotted but yield indistinguishable results. We find that the structure factor and the stiffness have converged to the same value for $L\geq6$ (see the inset).  The identification of a supersolid in QMC (bottom panel of Fig.~\ref{fig2}) therefore qualitatively agrees with the same identification in MFT (top panel) for weak disorder.

Our parameter choices so far do not offer strong evidence for an intermediate disordered phase between the solid and the supersolid.  Starting from the incompressible solid phase the theorem of inclusions \cite{pollet:2009} implies a Griffiths-type scenario as we drive the system to a compressible state in infinite system sizes and at zero temperature.  For the parameters considered here we expect that the disordered solid phase plays a role analogous to the Bose-glass in the disordered BH model.  Thus the disordered solid should always, in principle, appear between the solid and the supersolid (dashed line in Fig.~\ref{fig1}). But this region remains very difficult to probe because large $V$ should considerably narrow the accessible parameter space for the disordered state.  A similar narrowing is observed for the large $U$ disordered BH model \cite{pollet:2009}.
More importantly, our MFT ignores quantum fluctuations and the QMC is done on finite sized systems.  We therefore expect very little evidence of the Griffiths-type region in our study.   Our methods (QMC and MFT) do accurately probe regions far from this Griffiths-type region.  We conclude that the supersolid is stable because it appears far from the transition.

\begin{figure}[t] 
   \centering
   \includegraphics[width=3.2in]{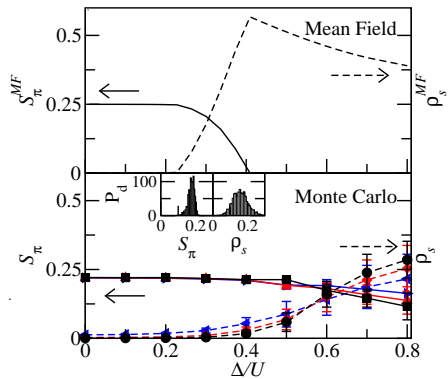} 
   \caption{(Color online) The top panel shows the structure factor (solid line: left $y$-axis) and stiffness (dashed line: right $y$-axis) vs. disorder strength for $\mu/U=0.4$, $zV/U=1$, $\beta U=20$, and $t/U=0.05$ computed using MFT.  The bottom panel shows the same but computed using QMC for $L=6$ (left and right triangles), $L=8$  (diamonds and up triangles), and $L=10$ (circles and squares) where the lines are guides to the eye.  Both methods imply that increasing just disorder strength can drive the solid into a supersolid.  The insets show distribution histograms ($P_{d}$) for $\Delta/U=0.6$ and $L=10$. }
   \label{fig3}
\end{figure}

We increase disorder to test the extent to which the solid (and supersolid) withstand spatial disorder in the chemical potential.  
To estimate the effects of increasing disorder on the solid phase we first increase $\Delta$ in our MFT.  The top panel of Fig.~\ref{fig3} tracks the structure factor and the stiffness as we increase disorder starting from the solid.  Here we see that disorder starts to trigger a supersolid near $\Delta/U\sim0.2$.  The theorem of inclusions requires that the solid becomes a disordered phase for $\Delta=E^{\text{S}}_{g}/2$ followed by another transition at $\Delta_{c}>E^{\text{S}}_{g}/2$.  Our MFT estimate for the onset of a supersolid at  $\Delta/U\sim0.2$ is consistent with this requirement because it is larger than a MFT estimate for the disappearance of the solid for $\Delta$ at $E^{\text{S}}_{g}/2U\approx 0.15$ with $t/U=0.05$.  

We construct a mean field site-percolation picture to understand the disorder induced supersolid on a SC lattice.   In the parameter regime considered here, the percolating supersolid is a percolating superfluid coexisting with a solid.  Consider a point in the solid phase on the $t$ vs. $\mu$ phase diagram near the solid-supersolid transition in the $\Delta=0$ limit.  MFT treats each site individually.  Increasing disorder will tend to move individual sites into a state nearby in phase diagram, i.e., the supersolid, if possible.  For a uniform probability distribution of chemical potentials in the interval $[\mu-\Delta,\mu+\Delta]$ a single site will have an energy in a specified range $[\mu+\epsilon_{1},\mu+\epsilon_{2}]$ with a probability $P(\epsilon_{1} \leq \epsilon\leq \epsilon_{2} )=(\epsilon_{2}-\epsilon_{1})/2\Delta$.  Using the meanfield boundaries of the solid phase and the site percolation threshold for a SC lattice ($p_{c}=0.31$ \cite{isichenko:1992}), we estimate a critical disorder strength for percolation of the supersolid through the solid, $\approx0.13$.  This estimate is consistent with QMC calculations presented below.  Percolation thus provides an intuitive picture for the onset of a supersolid with increasing disorder in SC lattices.   We note that the non-zero stiffness and structure factor completely distinguish the percolating supersolid from the Bose glass and a percolating superfluid.

\begin{figure}[t] 
   \centering
   \includegraphics[width=2.5in]{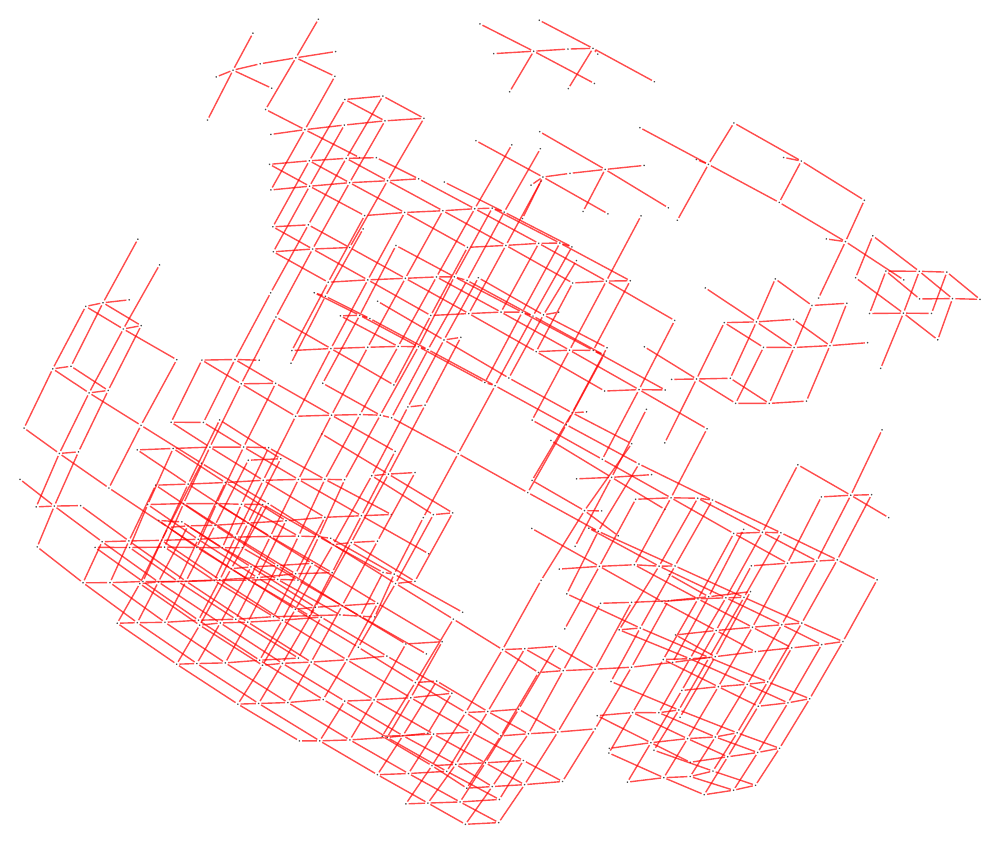} 
   \caption{(Color online) QMC results for a single disorder configuration for the parameters of Fig.~\ref{fig3} with $L=10$ and $\Delta/U=0.6$ but for $\beta U=50$.  We find solid order among all sites but we only plot a site $i$ if it has large density fluctuations: $\langle n_{i}^{2}\rangle-\langle n_{i}\rangle^{2}>0.2$.  We draw all available bonds from $i$ within the cube.  The supersolid is allowed to percolate from one edge to another only if local density fluctuations permit local superfluid order (the off-diagonal matrix elements in Eq.~(\ref{stiffness})) along enough connected bonds.     }
   \label{fig4}
\end{figure}

We use QMC to include quantum fluctuations as we increase disorder starting from the solid phase.  Figures~\ref{fig3} (bottom panel) and \ref{fig4} show QMC results suggesting that quantum fluctuations allow disorder to trigger a percolating supersolid from a solid.  Large disorder strengths limit QMC calculations.  For example, convergence for the data point at $\Delta/U=0.8$ and $L=10$ in Fig.~\ref{fig3} required $\sim10^{4}$ CPU hours on R410 PowerEdge Dell servers. 

The top and bottom panels of Fig.~\ref{fig3} are qualitatively consistent.  In the bottom panel we see that as we increase $\Delta$: i) The solid remains robust for $\Delta/U\lesssim 0.2$.  ii) The stiffness has a slight, $L$ dependent upturn for $0.2 \lesssim \Delta/U\lesssim 0.4$.  Here we expect a Griffiths-type phase, the disordered solid, in the thermodynamic limit and $\beta\rightarrow\infty$.  iii) The supersolid begins to percolate through the lattice for $0.4 \lesssim \Delta/U\lesssim 0.6$ where $L$ independent data support nonzero values for both $S_{\pi}$ and $\rho_{S}$.  iv) A weak supersolid for $\Delta/U\gtrsim 0.6$ (where the percolating supersolid order parameters may have $L$ dependence) and eventually a Bose-glass at large $\Delta$.  

We have established a platform for the study of disordered supersolids.  Our MFT and QMC results both show that supersolids are stable against weak spatial disorder on the SC lattice.  The supersolid remains stable far from expected Griffiths-type regions in parameter space.  We also find a striking transformation of the solid into a supersolid with increasing disorder strength.  Further work will explore critical properties and map out the full QMC phase diagram.   

We thank N. Prokof'ev and M. Troyer for valuable discussions.  
We acknowledge support from the Jeffress Memorial Trust (J-992), AFOSR (FA9550-11-1-0313), and DARPA-YFA (N66001-11-1-4122).


\begin{thebibliography}{99}

\bibitem{matsubara:1956} T. Matsubara and H. Matsuda,
Prog. Theor. Phys. {\bf 16}, 569 (1956).

\bibitem{liu:1973} K. Liu  and M.E.  Fisher, 
J. of Low Temp. Phys. {\bf 10}, 655 (1973).

\bibitem{otterlo:1995} A. van Otterlo \emph{et al.},
Phys. Rev. B {\bf 52}, 16176 (1995).

\bibitem{micnas:1990}  R. Micnas \emph{et al.},
Rev. Mod. Phys. {\bf 62}, 113 (1990). 

\bibitem{kim:2004} E. Kim and M. H. W. Chan, 
Nature {\bf 427}, 225 (2004).

\bibitem{clark:2007} A. C. Clark \emph{et al.}, 
Phys. Rev. Lett. {\bf 99}, 135302 (2007).

\bibitem{scalettar:1995} R.T. Scalettar \emph{et al.},
Phys. Rev. B {\bf 51}, 8467 (1995).

\bibitem{batrouni:2000} G.G. Batrouni \emph{et al.}, 
Phys. Rev. Lett. {\bf 84}, 1599 (2000).

\bibitem{sengupta:2005} P. Sengupta \emph{et al.},
Phys. Rev. Lett. {\bf 94}, 207202 (2005).

\bibitem{bernardet:2002}  K. Bernardet \emph{et al.},
Phys. Rev. B {\bf 66}, 054520 (2002).

\bibitem{imry:1975} Y. Imry and S. Ma,
Phys. Rev. Lett. {\bf 35}, 1399 (1975).

\bibitem{yamamoto:2009}  K. Yamamoto \emph{et al.},
Phys. Rev. B {\bf 79}, 094503 (2009).

\bibitem{fisher:1989} M.P.A. Fisher \emph{et al.}, 
Phys. Rev. B {\bf 40}, 546 (1989).
 
\bibitem{pollet:2009} L. Pollet \emph{et al.}, 
Phys. Rev. Lett. {\bf 103}, 140402 (2009);
V. Gurarie \emph{et al.}, 
Phys. Rev. B {\bf 80}, 214519 (2009). 

\bibitem{chen:2008} Y.P. Chen \emph{et al.}, 
Phys. Rev. A {\bf 77}, 033632 (2008).

\bibitem{billy:2008} J. Billy  \emph{et al.},
 Nature {\bf 453}, 891 (2008).

\bibitem{roati:2008} G. Roati \emph{et al.},
Nature {\bf 453}, 895 (2008).

\bibitem{pasienski:2010} M. Pasienski \emph{et al.}, 
Nature Phys. {\bf 6},  677 (2010).

\bibitem{goral:2002} K. Goral \emph{et al.}, 
Phys. Rev. Lett. {\bf 88}, 170406 (2002).

\bibitem{scarola:2005} V. W. Scarola and S. Das Sarma, 
Phys. Rev. Lett. {\bf 95}, 033003 (2005). 

\bibitem{bloch:2008} I. Bloch \emph{et al.},
Rev. Mod. Phys. {\bf 80}, 885 (2008). 

\bibitem{krauth:1991}  W. Krauth \emph{et al.},
Phys. Rev. Lett. {\bf 67}, 2307 (1991).

\bibitem{sheshadri:1995} K. Sheshadri \emph{et al.},
Phys. Rev. Lett. {\bf 75}, 4075 (1995).

\bibitem{dang:2009} L. Dang \emph{et al.},
 Phys. Rev. B {\bf 79}, 214529 (2009).

\bibitem{pollock:1987} E. L. Pollock and D. M. Ceperley,
Phys. Rev. B {\bf 36}, 8343 (1987).

\bibitem{syljuasen:2002} A. W. Sandvik, Phys. Rev. B {\bf 59}, R14157 (1999); 
O.F. Syljuasen and A. W. Sandvik, Phys. Rev. E {\bf 66}, 046701 (2002).

\bibitem{albuquerque:2008} A. F. Albuquerque \emph{et al.},
J. of Mag. and Mag. Mat.  {\bf 310}, 1187 (2007); 
B. Bauer \emph{et al.}, 
J. Stat. Mech. (2011) P05001.

\bibitem{isichenko:1992} M. B. Isichenko, Rev. Mod. Phys. {\bf 64}, 961 (1992).

\end{thebibliography}
\end{document}